\documentclass[conference]{IEEEtran}

\usepackage{cite}      
\begin{document}

\title{Threshold-Controlled Global Cascading in Wireless Sensor Networks}

\author{\authorblockN{Qiming Lu and G. Korniss}
\authorblockA{Department of Physics,
Rensselaer\\
Troy, New York 12180--3590\\
Email: \{korniss,luq2\}@rpi.edu}
\and
\authorblockN{Boleslaw K. Szymanski}
\authorblockA{Department of Computer Science,
Rensselaer\\
Troy, New York 12180--3590\\
Email: szymab@rpi.edu}}

\maketitle

\begin{abstract}
We investigate cascade dynamics in threshold-controlled
(multiplex) propagation on random geometric networks. We find that
such local dynamics can serve as an efficient, robust, and
reliable prototypical activation protocol in sensor networks in
responding to various alarm scenarios. We also consider the same
dynamics on a modified network by adding a few long-range
communication links, resulting in a small-world network. We find
that such construction can further enhance and optimize the speed
of the network's response, while keeping energy consumption at a
manageable level.
\end{abstract}


%
\IEEEpeerreviewmaketitle

\section{Introduction}

Many of the modern and important technological, information and
infrastructure systems can be viewed as complex networks with a
large number of components \cite{BarabREV,MendesREV,NEWMAN_SIAM}.
The network consists of nodes (or agents) and (physical or
logical) links connecting the nodes. These links facilitate some
form of interaction or dynamics between the nodes. Spreading
information fast across such networks with efficient and
autonomous control is a challenging task. Wireless Sensor Networks (WSNs)
provide an example where understanding dynamical processes {\em on} the
network is crucial to develop efficient protocols for autonomous
operation.

A sensor network is comprised of a large number of sensor nodes
which monitor, sense, and collect data from a target domain and then
process and transmit the information to the specific sites (e.g.,
headquarters, disaster control centers). There are many potential
applications of sensor networks including military, environment and
health areas (for a taxonomy of sensors networks
see~\cite{mobile02}). There are fundamental differences between a
sensor network and other wireless ad-hoc networks. First, sensor
nodes are often densely deployed (typically 20 sensor per cubic
meter)~\cite{computer-networks02} so that the underlying network has
high redundancy for sensing and communications. Accordingly, the
size of sensor networks may be several orders of magnitude larger
than the other ad-hoc networks. Hence, scalability of sensor network
operations is of utmost importance (see, for
example~\cite{mobicom99} for scalable coordination challenges and
solutions or~\cite{book-bksby04} scalable, self-organizing designs
for sensor networks and for limits on achievable capacity and delay
in mobile wireless networks). Second, sensor nodes have limited
battery power without recharging capabilities. Nodes running out of
power may cause topology changes in sensor networks even without
mobility (see for example ~\cite{wman05} for scalable and
fault-tolerant routing and~\cite{icn05} for communal routing in
which some of the nodes take over routing for the sleeping
neighbors). Third, new sensors with fresh batteries may be injected
to a sensor network, already in use, to enhance and ensure its
correct operation. Finally, the sensor nodes may be deployed in
adversarial environments such as battlefields, hostile territories
or hazardous domains that make their management, control and
security very difficult. Combined with diverse environments, ranging
from deserts to rain forests, from urban areas to battlefields and
habitats of protected species~\cite{comm-magazine02}, these
challenges make designing sensor networks that can operate reliably
and autonomously (totally unattended) very difficult.

The focus of this paper is on outliers detection in wireless
sensor network but the challenge is the same as in the above mentioned work,
to design scalable, energy efficient algorithms for communication and coordination.
Outlier detection is an essential step which precedes most any analysis
of data. It is used either with the intention of suppressing the outliers
or amplifying them. The first usage (also known as data cleansing)
is important when the analysis carried on the data is not robust.
Examples for such applications are optimization tasks, including routing
(where erroneous data may lead to infinite loops). The second usage
is important when looking for rare patterns. This often happens in
adversarial domains such as battlefield monitoring, controlling a boundary or
a perimeter of protected objects or intrusion detection.

Outliers are caused not only by external factors, but also by imperfections
in the acquisition of the data.
They typify error prone systems, specifically those which ought to
operate in harsh environmental conditions and make imperfect measurements
of external phenomena. Another setting in which outliers may occur
is whenever an adversary can control the measurement (but not the
computation and communication) of a device. In this setting, outliers
detection can either detect the manipulation of the data, or limit
the extent to which the data is manipulated. Thus, in some settings,
outliers detection limits the ability of an adversary to divert the
result.

Several factors make WSNs especially prone to
difficulties in outliers detection.
First, WSNs collect their data from the real world using imperfect sensing
devices. Next, they are battery operated and thus their performance
tend to deteriorate as power is exhausted. Moreover, as sensor networks
may include thousands of devices, the chance of error accumulates in them
to high levels. Finally, sensors are especially exposed to manipulation by
adversaries in their usage for security and military purposes. Hence, it is
clear that outlier detection should be an inseparable part of any
data processing that takes place in sensor networks.
In this paper, we investigate a simple model of cleansing and amplifying outliers in wireless
sensor networks. WSN environments pose several restrictions
on outliser computation, such as: (i) it has to be done in-network because
communication of raw data would deplete batteries, (ii) communication may
often be asymmetric, (iii) the data is streaming, or at least dynamically
updated, and (iv) both spatial and temporal locality of
the data are important for the result -- data points sampled by nearby
sensors during a short period of time ought to be more similar than
ones sampled by far off sensors over a large time interval. Hence,
in this paper, we assume that the number of nodes reporting outliers is
significant in making a decision to amplify or not the outlier discovery.

Sensor networks are both spatial and random. As a large number of
sensor nodes are deployed, e.g.,  from vehicles or aircrafts, they are
essentially scattered randomly across large spatially extended
regions. In the corresponding abstract graph two nodes are
connected if they mutually fall within each others transmission
range, depending on the emitting power, the attenuation function
and the required minimum signal to noise ratio. Random geometric
graphs (also referred to as Poisson/Boolean spatial graphs),
capturing the above scenario, are a common and well established
starting point to study the structural properties of sensor
network, directly related to coverage, connectivity, and
interference. Further, most structural properties of these
networks are discussed in the literature in the context of
continuum percolation \cite{percolation,penrose,Dall_2002}.

The common design challenge of these networks is to find the
optimal connectivity for the nodes: If the connectivity of the
nodes is too low, the coverage is poor and sporadic. If the node
connectivity is too high, interference effects will dominate and
results in degraded signal reception
\cite{Kumar2000,Kumar2004,phtr_networks,krause04}. From a
topological viewpoint, these networks are, hence, designed to
``live" somewhat above the percolation threshold. This can be
achieved by adjusting the density of sensor nodes and controlling
the emitting power of the nodes; various power-control schemes
have been studied along these lines \cite{Kumar2000,krause04}. In this paper
we consider random geometric graphs above the percolation threshold, as minimal models
for the underlying network communication topology. The focus of this work is to study
novel cascade-like local communication dynamics {\em on} these well studied graphs.

Here, we focus on the scenario where the agents (the individual
sensors) are initially in an {\em inactive} mode, typically
performing some periodic local measurements. However, if an
alarm-triggering event is detected locally by a (few) agent(s),
the network, as a whole should ``wake-up'' (all agents turning to
an active state), to closely monitor the spatial and temporal
behavior of the underlying phenomena that caused the alarm (e.g.,
spread of a fire or toxic chemicals). This process of turning
agents from an inactive state to an active one, requires some kind
of local rules between agents. There are three (somewhat
conflicting) objectives for constructing an optimal protocol:
\begin{enumerate}
\item reliability, so local erroneous events or
  false-alarms are suppressed and do not result in a ``global wake-up'';
\item speed, so that sensors can monitor the
  underlying physical, chemical, etc. phenomena;
\item energy efficient, so main concern in sensor networks, namely energy
limitation is addressed.
\end{enumerate}
To this end, we will consider a simple threshold-based model
(or multiplex propagation)
\cite{Grano78,WATTS_2002,CME_2005} on the sensor network with the potential to
efficiently facilitate the transition of the nodes from an inactive to an active state.
First, we will consider the threshold-based cascade dynamics on random
geometric networks.
Then we will experiment with the ``addition'' of a few long-range
communication links, representing multi-hop transmissions. In
particular, we will investigate the benefits in shortening the global
transition time versus the increase in communication (and therefore also
energy) costs. Such
networks, commonly referred to as small-world networks
\cite{Watts98,Watts99},
has long been known to speed up the spread of local information to
global scales \cite{BarabREV,MendesREV,NEWMAN_SIAM,Watts98,Watts99}
and to facilitate autonomous synchronization in coupled
multi-component systems \cite{STROG_NTWK_REV,BARAHONA02,KNGTR03a,KHK_PRL04,KHK_PRL05}.

\hfill

\section{Threshold-Based Propagation on Random Geometric Networks}
\subsection{Random Geometric Networks}

As mentioned in the Introduction, in this paper we consider random
geometric graphs \cite{percolation,penrose,Dall_2002} as the
simplest topological structures capturing the essential features
of ad hoc sensor networks. $N$ nodes are distributed at uniformly
random in an $L\times L$ spatial area. For simplicity we consider
identical radio range $R$ for all nodes. Two nodes are connected
if they fall within each other's range. An important parameter in
the resulting random geometric graph is the average degree or
connectivity $\alpha$ (defined as the average number of neighbors
per node $\overline{k}$),
$\alpha\equiv\overline{k} = 2K/N$,
where $K$ is the total number of links and $N$ is the number of
nodes. In random geometrical networks, there is a critical value
of the average degree, $\alpha_c$, above which the largest
connected component of the network becomes proportional to the
total number of nodes (the emergence of the giant component)
\cite{percolation,penrose,Dall_2002}. For a given density of nodes
$\rho$, there is a direct correspondence between the degree of
connectivity $\alpha$ and the radio range $R$ of each node
\cite{percolation,penrose,Dall_2002},
\begin{equation}
\alpha =\rho \pi R^2\;.
\label{degree_R}
\end{equation}
In what follows, for convenience, we will use $R$ instead of $\alpha$, as the relevant
parameter controlling the connectivity of the network.

\subsection{Threshold-Controlled Propagation}
The phenomenon of large cascades triggered by small initial
shocks, originally motivated by propagation in social networks
\cite{Grano78}, can be described by a simple {\em threshold-based}
model \cite{Grano78,WATTS_2002,CME_2005}. This model considers the
dynamics {\em on} a network of interacting agents (wireless
sensors in the present context), each of which must decide between
two alternative actions and whose decisions depend explicitly on
the actions of their neighbors according to a simple threshold
rule. Unlike in simple diffusive propagation, such as the spread
of a disease, where a single node is sufficient to ``infect''
(activate) its neighbors, in threshold-based (multiplex)
propagation node activation requires simultaneous exposure to
multiple active neighbors. Here, we implemented these simple rules
for random geometric networks, capturing the topological features
of wireless sensor networks.

The detailed description of the threshold model is as follows.
Each agent can be in one of two states: state $0$ or state $1$,
corresponding to the agent being inactive or active, respectively.
Upon observing the states of its $k$ neighbors, an agent turns its
state from 0 to 1 only if the fraction of its active neighbors is
equal to or larger than a specific threshold $\varphi$. In this
work we are interested in the temporal characteristics of global
cascades (network ``wake-ups''), hence once a node turns active,
it remains active for the duration of the evolution of the system.

We consider a system of $N$ agents located at the nodes of a
random geometric network. Each agent is characterized by a fixed
threshold $0 \le \varphi \le 1$. For simplicity all agent have the
same threshold. Initially the agents are all off (in state 0). The
network is perturbed at time $t=0$ by a small fraction of nodes
that are switched on (switched to state 1). The number of active
nodes then evolves at successive time steps with all nodes
updating their states simultaneously (synchronous updating) or in
random, asynchronous order (asynchronous updating) according to
the threshold rule above. Once a node has switched on, it remains
on (active) for the duration of the experiment.

For sensor networks, we are interested in the behavior of the
network under emergent situation which is represented by small
perturbation in the initial condition. In our investigation,
the focus is on:
\begin{enumerate}
\item the probability that a successful
\textit{global cascade} will be ignited by small fraction of initial
seed(s);
\item time needed for a global cascade, that is how
fast an initial shock will spread out to the entire network; and
\item the energy used for communication between agents in a
successful global cascade.
\end{enumerate}
The last quantity is an important factor in designing wireless
sensor networks. Here the term \textit{cascade} refers to an event
of any size triggered by initial seed(s), whereas \textit{global
cascade} is reserved for sufficiently large cascades
(corresponding to a final fraction of active agents, larger than a
cutoff fraction of large, but finite network).

\subsection{Simulation and Analysis}
We simulated systems consisting of $N=10^4$ sensor nodes distributed
randomly in a $L\times L =10^3\times 10^3$ (in arbitrary units) two-dimensional
region with periodic boundary conditions.
These nodes are employed (for simplicity) with identical communication range
$R$ to form a random geometric network.
For a fixed density of sensor nodes, $\rho=N/L^2$, the relationship
between the average degree $\alpha$ and the radio range $R$ is given by
Eq.~(\ref{degree_R}).
\begin{figure}[tb]
\vspace*{2.0truecm}
       \includegraphics{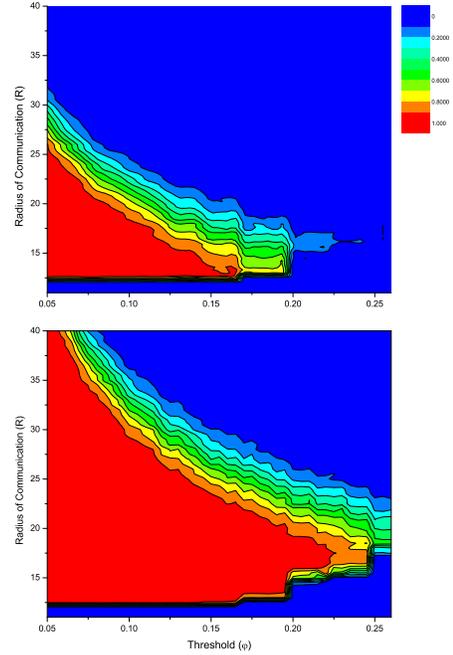}
\vspace*{7.0truecm} \caption{Phase diagram in the plane of
threshold and radio range $(\varphi,R)$. The cascade window is
enclosed by two different types of boundaries. (a) the cascade
window when the size of the initial seed is one. (b) the cascade
window when the size of the initial seed is three. Both graphs are
obtained at the system size $N=10^4$ and averaged over 1,000
simulation runs.}
\label{fig2}
\end{figure}

\begin{figure}[t]
\vspace*{2.0truecm}
       \includegraphics{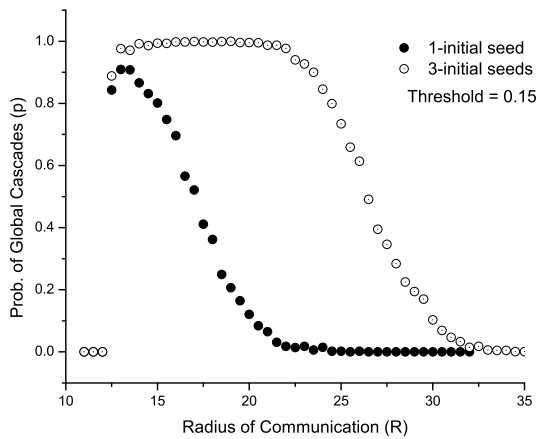}
\vspace*{4.0truecm} \caption{Cross section of cascade window from
Fig.~\ref{fig2}, at $\varphi=0.15$, with one initial seed node
(solid circles) and with three initial seed nodes (open circles).
Choosing any value at which these two curves separate sufficiently
far away from which other enables the network to suppress the global
cascade triggered by poorly supported events. The radius $R$ is
connected to the average network degree $\alpha$ according to Eq.
(1).} \label{fig3}
\end{figure}

Figure~\ref{fig2} displays the phase diagram for the cascade
dynamics on the $(\varphi,R)$ plane in terms of the probability of
global cascades for two differentinitial seed size. Each point in
the graph is obtained by averaging over $1000$ simulations
(including different network topologies and initial cascade
seeds). For seed size one, only a single node is activated
initially in the network. For seed size three, we randomly select
three neighboring (connected) nodes as a seed and turn them active
as the initial condition. In the phase diagram, by fixing the
threshold $\varphi$ and going along the line parallel to the
$R$-axis, hence increasing the radius of communication $R$, the
system exhibits two different phase transition as shown in
Fig.~\ref{fig3}. The first transition occurs at about
$R_{c1}\simeq12.5$ (corresponding to $\alpha\simeq 4.9$) where the
probability of global cascades sharply rises from 0 to around 1
for both initial seed sizes. We refer to this phase transition as
phase transition I. Further increasing the communication range
$R$, the probability of a global cascade slowly drops to 0. This
phase transition is referred to as phase transition II. Phase
transition I is inherently related to the emergence of the giant
component in a random geometrical network
\cite{percolation,penrose,Dall_2002}. Below the critical value
$R_{c1}$, the network is poorly connected, hence no cascade can
spread to global scales. Above $R_{c1}$ the giant component can
support global cascades, depending on the threshold $\varphi$.
This transition across $R_{c1}$ is sharp, related to the scaling
properties of the giant component of the random geometric network.
For sufficiently small threshold values $\varphi<1/\alpha$ the
probability of global cascades is close to 1 in a well connected
graph, yielding the upper boundary of the region where global
cascades are possible, $R_{c}(\varphi)$. As the range $R$ is
increased, while the threshold $\varphi$ is held fixed, nodes will
have have so many neighbors that they cannot be activated by a
single active agent. Since the relationship between the average
degree and the range follows Eq.~(\ref{degree_R}), the approximate
location of the boundary associated with phase transition II
scales as
\begin{equation}
R_{c}(\varphi)\sim 1/\sqrt{\varphi}\;.
\label{phtr2}
\end{equation}

\begin{figure}[t]
\vspace*{2.0truecm}
       \includegraphics{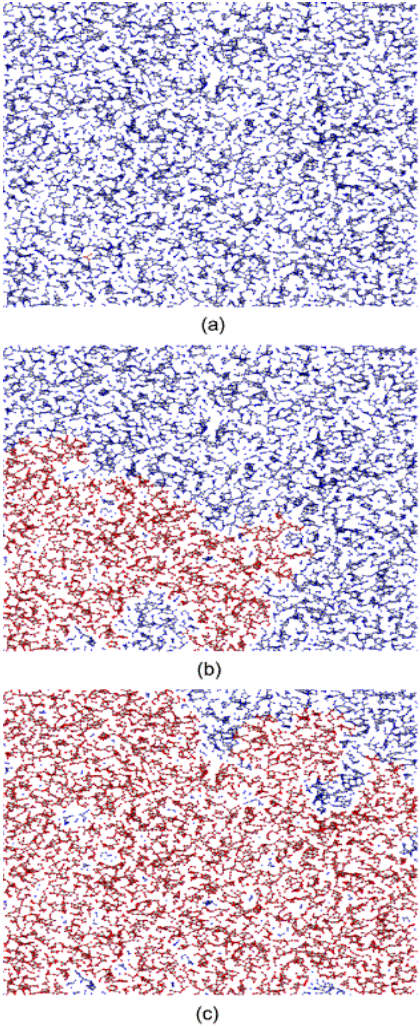}
\vspace*{13.5truecm} \caption{Snapshots of the regular random
geometric network evolving with time after the initial shock. Blue
dots are in state 0, red ones are in active state. Black lines are
local links. Snapshots are taken at time-step (a) t=0; (b) t=80; (c)
t=170. The network implemented the synchronous updating with
$N=10^4$, $\varphi=0.12$, and $R=16.0$.}
\label{snap-reg}
\end{figure}

Figure \ref{snap-reg} is the snapshots from a successful global
cascade with non-periodical boundary condition. At time $t=170$ the
fraction of active nodes exceeds 0.85 and completes the global
cascade.

In reality, it is possible that one of the sensors fails or turns
active and sends out an alarm message by accident. One of our
objectives is to construct local dynamics where global ``wake-ups''
due to erroneous events or miss-alarms are suppressed or possibly
excluded. This should be accomplished in an autonomous fashion,
i.e., without any external filtering or intervention. To this end,
first, it is reasonable to assume that the probability that a
specific node will send out an erroneous alarm is very small. Thus,
if, e.g., three neighboring nodes become active simultaneously, it
should be considered to be a real event. Comparing
Fig.~\ref{fig2}(a) (one initial seed node) and Fig.~\ref{fig2}(b)
(three initial seed nodes), it is clear that to prevent miss-alarms
by eliminating global cascades triggered by one initial seed, while
allowing them when triggered by the real event defined as above, we
should chose the region outside of the global cascades (blue/dark
colored region) of Fig.~\ref{fig2}(a) but inside the global cascade
region (red/light colored region) of Fig.~\ref{fig2}(b) by picking
proper $R$ and $\varphi$. Fig.~\ref{fig3} is the cross section of
phase diagram at $\varphi=0.15$. We can see that if the radius of
communication range is within $20<R<25$, the probability of global
cascades triggered by three initial seed nodes is still very close
to 1 while the probability of global cascades triggered by a single
initial seed essentially drops to 0. Hence, by choosing the proper
values of $R$ and $\varphi$, according to the phase diagrams, we can
prevent miss-alarms and, at the same time, ensure that real events
will trigger global cascades with near certainty.

\begin{figure}[t]
\vspace*{2.0truecm}
       \includegraphics{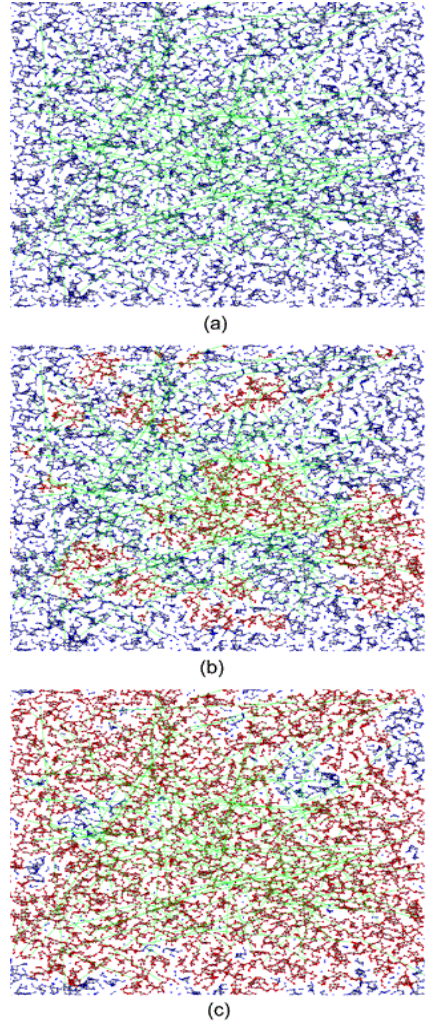}
\vspace*{13.5truecm} \caption{Snapshots of the small-world network
evolving with time after the initial shock. Green lines are random
long-ranged links. Snapshots are taken at time-step (a) t=0; (b)
t=28; (c) t=45. Messages propagate much faster than in regular
random geometric networks. The network implemented the synchronous
updating with $N=10^4$, $\varphi=0.12$, and $R=16.0$.}
\label{snap-sw}
\end{figure}

\section{Threshold-Model with Small-World Links}ea
The small world phenomenon originates from the observation that
individuals are often linked by a short chain of acquaintances.
Milgram \cite{Milgram} conducted a series of mail delivery
experiments and found that an average of ``six degrees of
separation'' exists between senders and receivers. The small-world
property (very short average path length between any pair of
nodes) were also observed in the context of the Internet and the
world wide web. Motivated by social networks \cite{Watts99},
and to understand network structures that exhibit low
degrees of separation, Watts and Strogatz \cite{Watts98} considered the
re-wiring of some fraction of the links on a regular graph, and observed that by
re-wiring just a small percentage of the links, the average path
length was reduced drastically (approaching that of random
graphs), while the clustering remains almost constant (similar to
that of regular graphs). This class of graphs was termed small-world
graphs to emphasize the importance of random links acting as
shortcuts that reduce the average path length in the graph. In the
following, we will continue to study random geometric graphs in
the context of wireless sensor networks but with the goal of
investigating the applicability of the small world concept to
these networks. The topological properties (such as the shortest
path and the clustering coefficient) of small-world-like sensor
networks have been studied in~\cite{HELMY_2003}. Here, we focus on the
effect of adding some random communication links between possibly
distant nodes, on {\em the dynamics} on the network.

\begin{figure}[t]
\vspace*{2.0truecm}
       \includegraphics{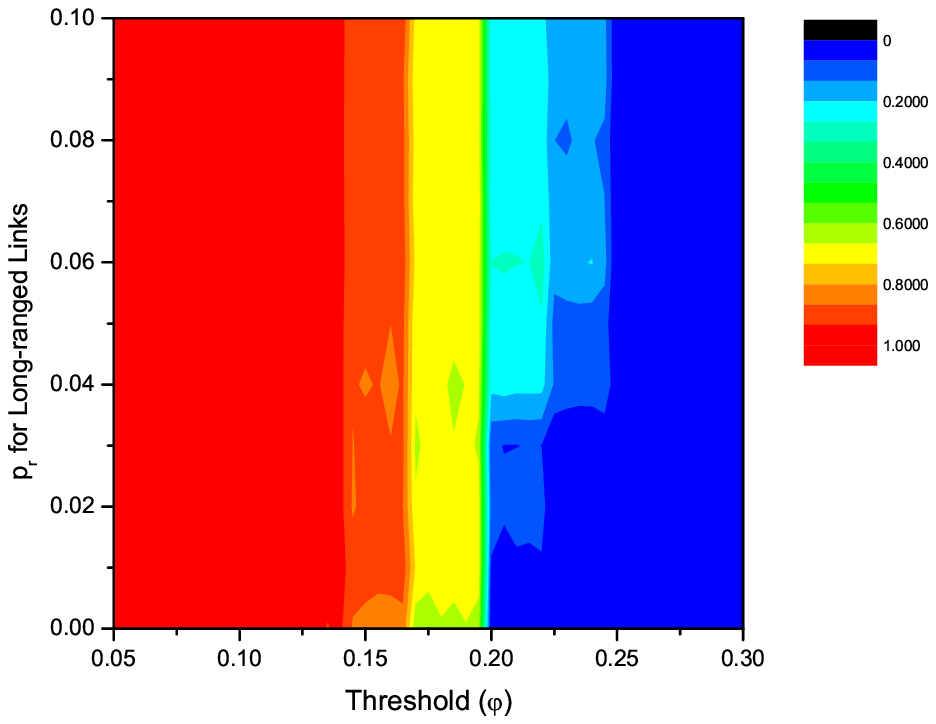}
\vspace*{4.0truecm} \caption{Phase diagram in the plane of
threshold and the probability of long-ranged links $(\varphi,p_r)$. The
cascade window is enlarged as we increase the probability of
long-ranged links ($p_r$). The graph is obtained at the system size
$N=10^4$ with $R=14$ and averaged over $1,000$ simulations.}
\label{fig4}
\end{figure}

As previously, we randomly deploy $N=10^4$ sensor nodes in a
$L\times L=10^3\times 10^3$ area, set a fixed radio range $R$ to
connect them, and generate the corresponding random geometric
network. In addition, we also add random ``long-range" links to
this backbone to construct a small-world-like network.
Random (possibly long-range) connections are constructed by {\em
adding} a fixed number of random links, so that the total number
of random added edges is $p_r N$ with $p_r\ll 1$. Alternatively,
one can construct statistically identical networks by adding a
random link emanating from each node with probability $p_r$. The
procedure has several different realizations, depending on how
$p_r$ (in the above probabilistic interpretation)
varies with the underlying spatial distance between the randomly connected two nodes.
\begin{enumerate}
\item $p_r=constant$, in which case there are no restrictions on the
length of random long range links;
\item $p_r\propto 1/d^\delta$, where $d$ is the distance between the two
chosen nodes and $\delta$ is a parameter;
\item $p_r=constant$ for $d\le d_c$ and $p_r=0$ for $d>d_c$, where $d$ is the
distance between the two picked nodes and $d_c$ is a parameter to which we
refer as \textit{cutoff distance}.
\end{enumerate}

The dynamics on the network is defined by the same threshold model that
was discussed
earlier. Here we show results for the case when the size of the initial seed is set to
one. Thus, the dynamics is controlled by threshold $\varphi$, the radio range
$R$, and the probability of a long-range links $p_r$.

The addition of small-world links is expected to speed up
propagation (reduce time for global cascades to complete) in the
region where global cascades are possible \cite{CME_2005}. Focusing
on the three quantities outlined earlier, the probability of global
cascades (yielding the phase diagram), the average global cascade
times, and energy costs, compared to the original random geometrical
network are discussed below. The measured observables are averaged
over 1,000 simulations. The time needed for global cascades is
averaged over all successful global cascades with the same model
parameters.

Snapshots of a successful global cascade in the small-world network
is shown in Fig.~\ref{snap-sw}. Comparing the snapshots from regular
random geometric networks, nodes in small-world networks can be
ignited by his neighbors in distance via long-ranged links added to
the network, thus expedite the speed of message propagation.

\begin{figure}[t]
\vspace*{2.0truecm}
       \includegraphics{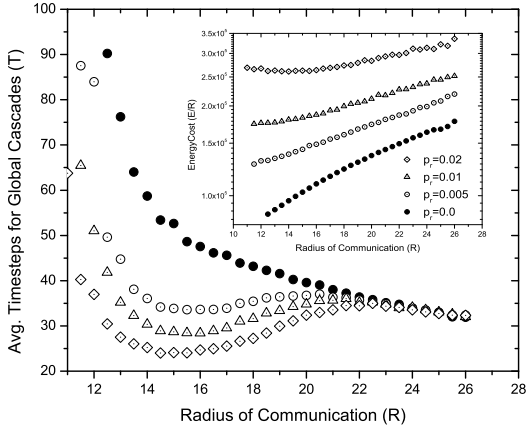}
\vspace*{4.0truecm} \caption{Effect of long-ranged links on the
dynamics of random geometric graph under threshold rule. The
introduction of long-ranged links decreased the time needed for
cascades but increased the energy cost. Symbols mean: $p_r=0$
(solid circles); $p_r=0.05$ (open circles); $p_r=0.01$ (open
triangles); $p_r=0.02$ (open diamonds). The graph is averaged over
$1,000$ simulations with synchronous updating and the threshold fixed at
$\varphi=0.12$.}
\label{fig5}
\end{figure}
\begin{figure}[t]
\vspace*{2.5truecm}
       \includegraphics{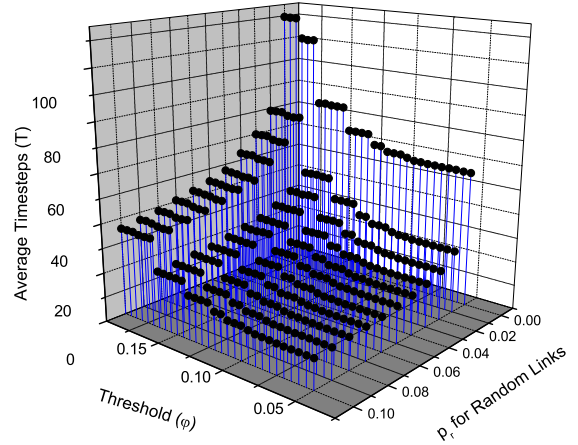}
\vspace*{4.0truecm} \caption{The reduction in global cascade time
resulting from the increased number of long-ranged links. Similar
behavior as in Fig.~\ref{fig5} is observed when radio range $R$ is
fixed. The graph is averaged over 1,000 simulations with synchronous
updating, $N=10^4$, and $R=14.0$.}
\label{fig6}
\end{figure}

Next, we define the energy consumption during a successful
global cascade. The total energy consumption in a wireless sensor
network is, in most part, due to communication between nodes,
computing, and storage (neglecting some smaller miscellaneous costs).
The communication is the main part of energy consumption,
and is directly related the network structure and dynamics, while energy used for
computing can be regarded as a constant, so it can be easily
evaluated. Consequently, in the following, we only consider the
energy consumption in communications between sensor nodes.

There are two kinds of communication in small-world graphs. First is
the local communication. According to the nature of sensor nodes,
the wireless broadcast is the most commonly used method for local
communication. Thus, the energy cost for local broadcast is
proportional to the square of radio range $R$,
$E_l = cR^2$,
where $c$ is a coefficient that we scaled to 1. It should also be
noted that a message send by a specific node can be received by all
his local neighbors, regardless of the number of them, by a single
broadcast. The second type of communication is the long-ranged one.
The energy cost for long-ranged communication depends on how this
communication is implemented, including solutions such as a
directional antenna, multi-hop transmission, global flooding, etc.
In this paper, we consider multi-hop transmissions in which messages
reach their destination following a multi-hop route. The energy cost
of such a transmission (for well-connected networks) can be approximated as
\begin{equation}
E_r \simeq cR^2(\frac{d}{R})=cRd=E_l\frac{d}{R},
\end{equation}
where $d$ is the distance between two nodes. Unlike in case of broadcast,
each pair of nodes that has a long-ranged link between them will
incur additional energy to communicate. Hence, the total energy consumption in
communication for a successful global cascade with $m$ local communications
and $n$ long-ranged communications is:
\begin{equation}
E=mE_l+nE_r
\end{equation}
Since $n=p_rN$, then denoting the average length of long-range links as
$\overline{d}$ and using an approximation $m\simeq N$ (the exact value of $m$
is equal to the number of nodes participating in the global cascade),
the energy cost can be rewritten as:
\begin{equation}
E \simeq
cN(R^2+p_r\overline{d}R)=NE_l\left(1+\frac{p_r\overline{d}}{R}\right)\;.
\label{total-energy}
\end{equation}
The first term corresponds to the local communication energy, while
the second term represents the additional energy needed for the
long-ranged links. The ratio of the long-ranged link communication energy to
the local communication energy is proportional to the  probability of
long-ranged links $p_r$ and the average link length and inversely proportional
to the radio radius $R$.
\begin{figure}[t]
\vspace*{2.5truecm}
       \includegraphics{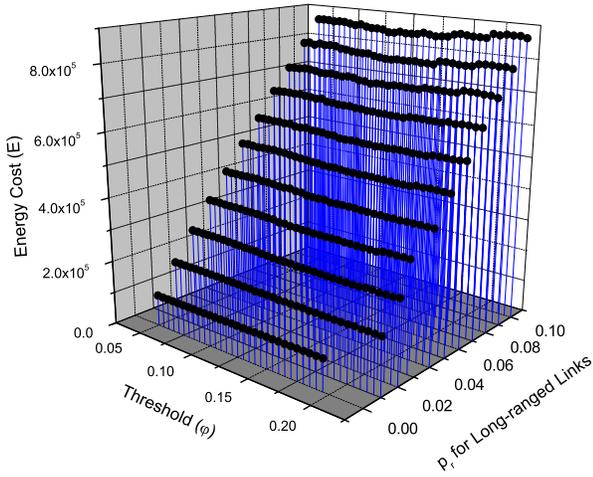}
\vspace*{4.0truecm}
\caption{The energy cost of  global cascades as a function of the threshold and the
density of long-ranged links.
The graph is averaged over 1,000 simulations with
synchronous updating, $N=10^4$, and $R=14.0$.}
\label{newfig}
\end{figure}

\begin{figure}[t]
\vspace*{2.0truecm}
       \includegraphics{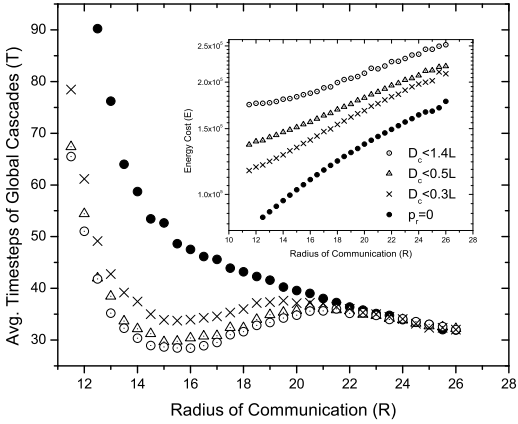}
\vspace*{4.0truecm} \caption{A comparison of average time and energy
costs with different cutoff distance of long-ranged links. Different
symbols are, $p_r=0$ (solid circles); $p_r=0.01$, $d_c=\sqrt{2}L$
(open circles); $p_r=0.01$, $d_c=0.5L$ (open triangles); $p_r=0.01$,
$d_c=0.3L$ (cross). The network implemented the synchronous updating
with the threshold fixed at $\varphi=0.12$.}
\label{fig7}
\end{figure}

Graphs for the cascade window, after adding the long-ranged links, are
similar to that of the model without long-range links [Fig.~\ref{fig2}].
The most significant difference is that the lower boundary of the cascade window
(which we referred to as phase transition I) shifts downwards by a small amount ($\sim0.5-1.0$)
toward smaller $R$. As we discussed above, a sufficiently large
value of $R$ is necessary for the entire graph to be well connected
and to be able to support global cascades.
By adding long-ranged links, it will be more likely that several small
clusters would be connected via these long-ranged links to form a
larger cluster, hence lowering the boundary of cascade windows.
Figure~\ref{fig4} displays the phase diagram in terms of
the probability of global cascades on the
plane of the threshold $\varphi$ and the probability of random long range
links $p_r$, $(\varphi, p_r)$, when the radio range $R$ is fixed at
14.0. Different from the results of Ref.~\cite{CME_2005} (as a
result of adding as opposed to re-wiring random links), the cascade
window somewhat enlarges at a specific region of $(\varphi, R)$ when
$p_r$ increases.

The advantage of the small-world links is that they can
significantly decrease the global cascade time. In other words,
alarms and messages propagate faster in small-world graphs than in
the original random geometric graphs, as shown in Figs.~\ref{fig5} and \ref{fig6}.
For fixed threshold, the time needed for a global cascade decreases
monotonously as $R$ increases in regular random geometric graphs.
In contrast, in small-world graphs, the average time is much lower
and reaches its optimal (minimum) value at $R\simeq15.0$. The more
long-ranged links are added to the network, the lower the average
time is. Meanwhile, due to the long-range communications, the
average energy cost for a successful global cascade is also
increasing, linearly with $p_r$, in agreement with
Eq.~(\ref{total-energy}) as can be seen in Fig.~\ref{newfig}.

\begin{figure}[t]
\vspace*{2.0truecm}
       \includegraphics{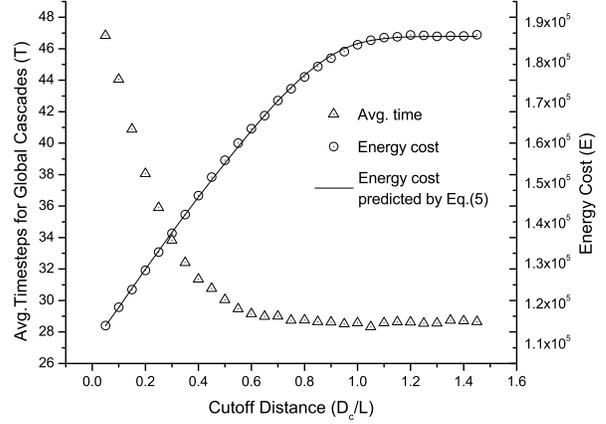}
\vspace*{4.0truecm} \caption{A dependence of the average time and
energy costs on the cutoff distance when the threshold and radio
range are both fixed. According to the shape of these two curves, we
can optimize the system behavior by varying the cutoff distance
$d_c$. Open triangles are the average time, open circles are the
energy cost and solid line is the energy cost predicted by
Eq.~(\ref{total-energy}) with $R=16$, $\varphi=0.12$ and synchronous
updating.}
\label{fig8}
\end{figure}

The above observations can be used to develop a scheme to compensate
the increase in energy cost caused by the added long-range links.
In the discussion above, the distribution of the distance of the long-range links was uniform.
Next, we explore the effects of ``suppressing" the occurrence of links with large spatial length.
While one can implement, e.g.,  a power-law or an
exponentially-tailed distribution for the spatial length of the added random links,
here, for simplicity and to ease technical realizations, we use a sharp cutoff for
$p_r(d)$,
\begin{equation}
p_r(d)=\left\{
    \begin{array}{rl}
        p_{r0} & \mbox{if $d\le d_c$} \\
        0 & \mbox{if $d>d_c$},
    \end{array}
\right.
\end{equation}
where $d_c$ is the cutoff distance for the long-ranged links. This
distance may represent the range of a uni-directional antenna of
special sensor nodes. Adding a small amount, $p_r N$, of such
random links to the network will provide
the long-range links in the network without the need to implement
multi-hop routing.

Scaling $d_c$ with the spatial size $L$ of the system, when
$d_c\ge \sqrt{2}L$, there is no restriction on the long-range link
length. The average time and energy costs for global cascade under
the restriction of long-range links is shown in Fig.~\ref{fig7}.

It can be seen that even putting a strong restriction of the longest
link distance ($d_c<0.5L$) only slightly increases the average time
for successful global cascades. However, under the same
restrictions, the energy cost drops significantly, proportionally to
the decrease in the average long-ranged link length, as predicted by
Eq.~(\ref{total-energy}). In particular, we
explored the behavior of the average time ($T$) and the energy cost ($E$)
vs. different cutoff distances ($d_c$) when $R$ and
$\varphi$ are both fixed [Fig.~\ref{fig8}].
The average time is close to its minimum when
$d_c>0.4L$ while the energy cost has still not saturated until
$d_c>1.0L$. Clearly, one can make a trade-off between the speed of
message propagating and the energy cost with different cutoff
distances of the long-range links. Depending on various applications
and scenarios, we can choose different $d_c$ to make the network
respond faster to emergencies or be more efficient in terms of
energy used.



%

\section{Conclusion}

In this paper, we explored cascade dynamics in threshold-controlled
(multiplex) propagation on random geometric networks as a simple yet
effective model of outliers cleansing and amplifying in wireless sensor
networks. Hence, the local dynamics of cascading can serve as an efficient,
robust, and reliable basic activation protocol for responding to various
alarm scenarios and distinguishing between false (few outliers indicating
alarm discovered) and real alarms (several outliers detected in close
proximity of each other). We also found that the network modified
by adding a few
long-range communication links, resulting in a small-world network, changes
the speed of the network's response. Hence,
such construction can further enhance and optimize the speed
of the network's response, while keeping energy consumption at a
manageable level.
More realistic ``backbone" topologies (beyond the minimalist random geometric graphs)
can be readily constructed and studied, based on the minimum requirements
of the local signal-to-noise for signal detection. Future studies will address
communication dynamics on such networks.

The presented research raised also several questions that we plan to
pursue in future research. One of the most fundamental questions is the impact
of synchrony or asynchrony of the alarm propagation. Another is about
the technical means of creating the long-range links. Instead using
multi-hop transmission, we can foresee mixing a small percentage of nodes with
directional antennas among all sensors and using those nodes to support
long-range links with just unit cost of communication. Such a realization
will improve the benefits of the scheme proposed in this paper.


\section*{Acknowledgment}
We thank Zolt\'an Toroczkai and Hasan Guclu for discussions
and sharing some of his (H.G.) earlier codes generating random geometric networks.
G.K. and Q.L. were supported in part by NSF Grant No.\ DMR-0426488 and B.K.S.
and Q.L. were supported in part by NSF Grant No.\ NGS-0103708.
This research was also supported in part by Rensselaer's Seed Program.



%

\end{document}